\documentclass{article}
\usepackage{colm2024_conference}

\usepackage{amsmath,amsfonts,bm}

\def\eqref#1{equation~\ref{#1}}

\def\1{\bm{1}}

\def\vmu{{\bm{\mu}}}

\def\va{{\bm{a}}}

\def\vh{{\bm{h}}}

\def\vs{{\bm{s}}}

\def\vv{{\bm{v}}}

\def\vx{{\bm{x}}}
\def\vy{{\bm{y}}}
\def\vz{{\bm{z}}}

\DeclareMathAlphabet{\mathsfit}{\encodingdefault}{\sfdefault}{m}{sl}
\SetMathAlphabet{\mathsfit}{bold}{\encodingdefault}{\sfdefault}{bx}{n}

\usepackage[T1]{fontenc}
\usepackage[utf8]{inputenc}
\usepackage{amsmath,amssymb}
\usepackage{graphicx}
\usepackage{booktabs}
\usepackage{multirow}
\usepackage{array}
\usepackage{xcolor}
\usepackage{microtype}
\usepackage{hyperref}
\usepackage{url}

\newcommand{\vP}{\bm{P}}
\newcommand{\vepsilon}{\bm{\epsilon}}
\newcommand{\vxi}{\bm{\xi}}
\newcommand{\cutetts}{CuteTTS}
\newcommand{\modelbase}{\cutetts{}}
\newcommand{\modeldistill}{\cutetts{}-distill}
\newcommand{\contar}{Cont. AR}
\newcommand{\contnar}{Cont. NAR}
\newcommand{\discar}{Disc. AR}
\newcommand{\hybridar}{Disc. AR + Cont. NAR}

\title{CuteTTS: Efficient and High-Quality Speech Synthesis via
Autoregressive Modeling of Continuous Latents}

\author{\normalfont
  Yuqian Zhang\textsuperscript{*,1,3},
  Yao Shi\textsuperscript{2},
  Kexin Huang\textsuperscript{3},
  Botian Jiang\textsuperscript{1,3},
  Zhe Xu\textsuperscript{1,3} \\
  Yiwei Zhao\textsuperscript{1,3},
  Min Liang\textsuperscript{2},
  Shuang Chen\textsuperscript{\(\dagger\),1,3},
  Xipeng Qiu\textsuperscript{1,3},
  Yu-Gang Jiang\textsuperscript{3}}

\hypersetup{
  pdftitle={CuteTTS: Efficient and High-Quality Speech Synthesis via Autoregressive Modeling of Continuous Latents},
  pdfauthor={Yuqian Zhang, Yao Shi, Kexin Huang, Botian Jiang, Zhe Xu, Yiwei Zhao, Min Liang, Shuang Chen, Xipeng Qiu, Yu-Gang Jiang}
}

\colmfinalcopy

\begin{document}
\maketitle

\begingroup
\renewcommand{\thefootnote}{}
\footnotetext{%
  \begin{tabular}{@{}l@{}}
    \textsuperscript{*} Work done during internship at OPPO.
    \href{mailto:yuqianzhang24@m.fudan.edu.cn}{\texttt{yuqianzhang24@m.fudan.edu.cn}} \\[-0.2ex]
    \textsuperscript{\(\dagger\)} Corresponding author:
    \href{mailto:chenshuang_fd@fudan.edu.cn}{\texttt{chenshuang\_fd@fudan.edu.cn}} \\[-0.2ex]
    \textsuperscript{1} Shanghai Innovation Institute \quad
    \textsuperscript{2} OPPO AI Center \quad
    \textsuperscript{3} Fudan University
\end{tabular}}
\endgroup

\begin{abstract}
  Zero-shot text-to-speech (TTS) now supports interactive assistants, personalized media, and accessibility tools. All TTS systems require faithful linguistic rendering, consistent speaker identity, and low-latency response. Yet compact streaming systems must preserve sufficient acoustic detail in a predictable low-rate latent sequence, while iterative diffusion sampling and classifier-free guidance multiply inference cost at every autoregressive step. To strike a balance between high-fidelity synthesis and low-latency inference, we present \textbf{CuteTTS}, a compact continuous-autoregressive TTS system. It combines semantically aligned causal VAE latents with patch-level autoregression, explicit speaker conditioning, and a bidirectional flow-matching head. We further introduce guidance--step distillation, which absorbs classifier-free guidance and multiple solver steps into a single interval-conditioned student. Evaluations on LibriSpeech and Seed-TTS-Eval demonstrate competitive intelligibility and speaker similarity in zero-shot voice cloning, while distillation lowers first-audio latency by 23.3\% and real-time factor by 40.8\% relative to the base model with comparable objective and subjective quality. These results provide a practical path toward continuous-autoregressive TTS that reconciles high-fidelity generation with the latency demands of real-time interaction. Code and checkpoint are available at \url{https://github.com/OPPO-Mente-Lab/CuteTTS}.
\end{abstract}

\section{Introduction}
\label{sec:introduction}

Zero-shot TTS enables conversational agents, personalized media, and assistive applications by reproducing the voice of an unseen speaker from a short reference utterance. Practical systems must preserve linguistic content and speaker identity while maintaining low latency. Existing approaches either model discrete codec tokens autoregressively~\citep{wang2023valle,kharitonov2023spear} or generate continuous speech representations through diffusion or flow matching~\citep{le2023voicebox,chen2025f5tts}. Hybrid streaming systems combine both paradigms~\citep{du2024cosyvoice,du2024cosyvoice2}. Despite substantial progress, achieving high-quality, speaker-faithful, and low-latency synthesis within a compact model remains challenging.

Continuous-autoregressive TTS retains the fidelity of continuous speech latents while supporting causal streaming generation~\citep{peng2025vibevoice,rouard2025calm}. Patch-based methods use a causal backbone to model dependencies across patches and a diffusion head to capture local correlations~\citep{jia2025ditar,zhou2025voxcpm}. However, compact streaming systems still face two challenges. Low-rate latents must balance reconstruction quality with autoregressive predictability, while iterative patch generation and LM-level classifier-free guidance require repeated model evaluations~\citep{ho2022cfg}. This computational overhead directly increases streaming latency.

In order to address these challenges, we present \cutetts, a compact continuous-autoregressive TTS system built around a low-rate causal speech representation, patch-level language modeling, and a flow-matching-based diffusion head. The representation is trained with a semantic alignment objective, and a separate speaker embedding directly conditions the TTS backbone. To accelerate the diffusion head, we train a single student on interval-averaged velocities along the CFG-guided teacher trajectory. The student therefore absorbs both the guidance branch and multiple numerical integration steps in one distillation objective.

Our main contributions are summarized as follows:
\begin{itemize}
  \item We develop \cutetts, an approximately $0.2$B-parameter streaming system that combines patch-level autoregression over continuous speech latents with a flow-matching diffusion head. It achieves competitive intelligibility and speaker similarity across LibriSpeech and Seed-TTS benchmarks, and is preferred to similarly sized baselines in subjective sound-quality comparisons.
  \item We show that semantic alignment of the low-rate speech representation improves intelligibility, while explicit speaker conditioning strengthens zero-shot voice cloning. Controlled studies of representation rate, patch size, and module-wise capacity scaling further clarify how temporal granularity affects reconstruction fidelity, autoregressive modeling difficulty, and local continuous generation.
  \item We introduce guidance--step distillation, which unifies classifier-free guidance removal and interval-wise integration in a single diffusion head. Under the paired efficiency protocol, the distilled model lowers mean first-audio latency by $23.3\%$ and mean real-time factor by $40.8\%$ while preserving synthesis quality close to that of the base model.
\end{itemize}

\section{Related Work}
\label{sec:related}

\paragraph{Discrete and continuous TTS.}
Codec language models cast TTS as autoregressive prediction over discrete speech
tokens~\citep{wang2023valle,kharitonov2023spear,peng2024voicecraft}.  Other
systems generate continuous-valued speech features in parallel using diffusion or
flow matching~\citep{popov2021gradtts,le2023voicebox,mehta2024matcha,
chen2025f5tts}.  Hybrid pipelines combine an autoregressive semantic model with a
separate continuous acoustic decoder, as in CosyVoice and CosyVoice~2
\citep{du2024cosyvoice,du2024cosyvoice2}.

\paragraph{Continuous autoregressive generation.}
Continuous-autoregressive models avoid a discrete codebook while
retaining causal sequence modeling.  MELLE directly predicts mel-spectrogram
frames~\citep{meng2025melle}; SALAD uses a per-token diffusion process
\citep{turetzky2024salad}; and DiTAR introduces patch-level generation with a
causal LM and a bidirectional diffusion decoder~\citep{jia2025ditar}.  Related
systems employ flow-based patch decoders, hierarchical continuous-latent
backbones, or consistency heads
\citep{wang2025felle,rouard2025calm,peng2025vibevoice,
an2025melatts,wu2025clear,zhou2026voxcpm2}.  \cutetts{} follows the patch-based
factorization of this family and focuses on efficient guided sampling.

\paragraph{Audio representation.}
Neural audio codecs obtain compact representations through quantization, trading
reconstruction fidelity against bitrate~\citep{defossez2023encodec}.  Continuous
latents avoid codebook discretization and can preserve fine-grained speech
variation, but a representation optimized only for reconstruction need not be
well suited to low-rate autoregressive prediction.  Semantic-VAE encourages VAE
latents to align with self-supervised speech features~\citep{niu2025semanticvae,
chen2022wavlm}, while MELA-TTS studies related representation alignment for
speech synthesis~\citep{an2025melatts}.  Motivated by these works, we use a
semantically aligned continuous audio representation for low-rate autoregressive
modeling.

\paragraph{Guidance and few-step sampling.}
CFG combines conditional and unconditional predictions to strengthen conditional
generation~\citep{ho2022cfg}.  Guided-model distillation can absorb this two-branch
computation~\citep{meng2023guideddistillation}, while progressive distillation and
consistency models reduce the number of sampling steps
\citep{salimans2022progressive,song2023consistency}.  More recent work learns
average velocities or conditions a model on the integration interval
\citep{geng2025meanflow,frans2025shortcut,wang2025intmeanflow}.  Our method builds
on these ideas but targets their joint effect within the diffusion head of a
continuous-autoregressive TTS model: the student learns the CFG-guided trajectory
and its interval-wise integration at the same time.

\section{CuteTTS}
\label{sec:method}

\begin{figure}[t]
  \centering
  \includegraphics[width=\linewidth]{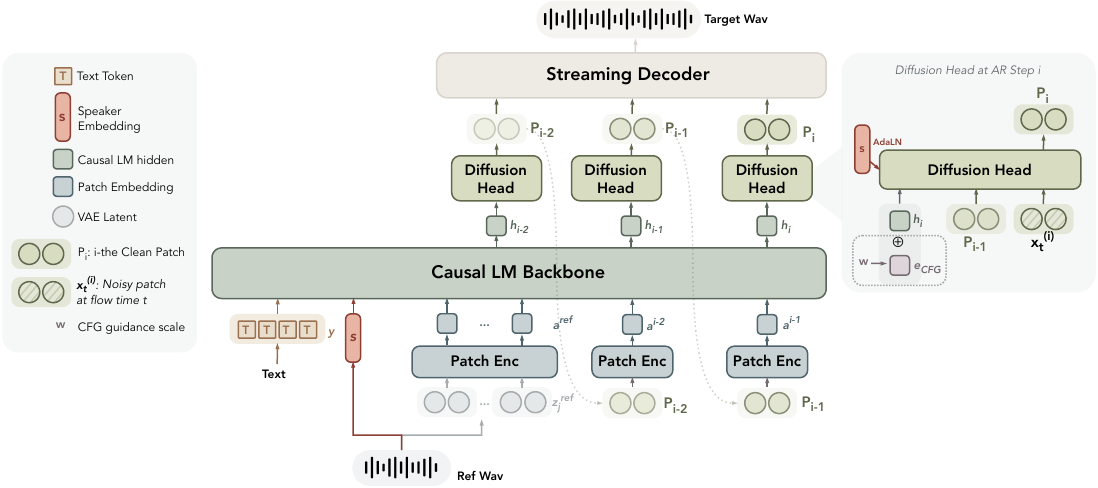}
  \caption{Overview of \cutetts. A patch encoder aggregates previously
  generated latent patches for the causal backbone. The final-layer LM hidden
  state at the current autoregressive position, preceding latent patch, and
  speaker embedding condition the diffusion head to generate the next
  continuous patch; for the distilled model, CFG weight $w$ is supplied as an
  additional condition. A causal VAE decoder converts completed patches to
  audio incrementally.}
  \label{fig:architecture}
\end{figure}

Given text $\vy$ and a reference utterance $\vx^{\mathrm{ref}}$,
\cutetts{} autoregressively generates a sequence of continuous latent patches
and decodes them into waveform chunks.  As illustrated in
Figure~\ref{fig:architecture}, the system consists of a causal audio VAE, a
patch encoder, a causal language-model backbone, a diffusion head, and a speaker
encoder.

\subsection{Causal audio VAE with semantic alignment}
\label{sec:vae}

We separately train a causal audio VAE that maps a 24-kHz waveform $\vx$ to continuous latents at 12.5~Hz and reconstructs $\hat{\vx}$.  Following \citet{kumar2023dac} and \citet{zhou2025voxcpm}, its encoder and
decoder use a DAC-style fully convolutional architecture with stacked causal convolutions and residual blocks supporting incremental encoding and decoding.

In order to avoid posterior variance collapse in continuous autoregressive modeling, we adopt a $\sigma$-VAE~\citep{sun2024latentlm,peng2025vibevoice}.  The encoder predicts a posterior mean
$\vmu=E_{\phi}(\vx)$, while the scale is sampled from a prescribed distribution
$p_{\sigma}$ rather than estimated from the input:
\begin{equation}
  \sigma\sim p_{\sigma}, \qquad
  \vz=\vmu+\sigma\vepsilon, \qquad
  \vepsilon\sim\mathcal{N}(\mathbf{0},\mathbf{I}).
  \label{eq:sigma_vae}
\end{equation}

Reconstruction fidelity alone does not ensure that a low-rate representation is
well suited to text-conditioned generation.  Inspired by
Semantic-VAE~\citep{niu2025semanticvae}, we align the latent sequence with a
frozen WavLM teacher~\citep{chen2022wavlm}.  The teacher features are projected
to the VAE dimension and temporally aligned to the latent rate, and a cosine loss
encourages the two representations to encode shared linguistic structure.

The VAE objective combines multi-resolution Mel-spectrogram reconstruction,
adversarial and feature-matching losses from multi-period and multi-resolution
discriminators, a lightly weighted KL regularizer, and semantic alignment:
\begin{equation}
  \mathcal{L}_{\mathrm{VAE}}
  = \lambda_{\mathrm{mel}}\mathcal{L}_{\mathrm{mel}}(\vx,\hat{\vx})
  + \lambda_{\mathrm{adv}}\mathcal{L}_{\mathrm{adv}}(\hat{\vx})
  + \lambda_{\mathrm{feat}}\mathcal{L}_{\mathrm{feat}}(\vx,\hat{\vx})
  + \lambda_{\mathrm{KL}}\mathcal{L}_{\mathrm{KL}}
  + \lambda_{\mathrm{sem}}\mathcal{L}_{\mathrm{sem}}.
  \label{eq:vae_objective}
\end{equation}
where $\mathcal{L}_{\mathrm{sem}}$ is the negative cosine similarity between the
latent sequence and the aligned teacher features.  After VAE training, only the
causal encoder and decoder are retained; the TTS model operates entirely in the
continuous VAE space.

\subsection{Explicit speaker conditioning}
\label{sec:speaker}

We train a compact ECAPA-style speaker encoder~\citep{desplanques2020ecapa} by
distilling a frozen WavLM Large speaker-verification
teacher~\citep{chen2022wavlm}.  Given a reference utterance
$\vx^{\mathrm{ref}}$, the student and teacher embeddings are
$\vs_{\mathrm{student}}=E_{\mathrm{student}}(\vx^{\mathrm{ref}})$ and
$\vs_{\mathrm{teacher}}=E_{\mathrm{teacher}}(\vx^{\mathrm{ref}})$,
respectively.  Both embeddings are $L_2$-normalized.  The student is optimized
with
\begin{equation}
  \begin{aligned}
  \mathcal{L}_{\mathrm{spk}}
  ={}&\mathcal{L}_{\mathrm{cos}}
  (\vs_{\mathrm{student}},\vs_{\mathrm{teacher}})\\
  &+\lambda_{\mathrm{pair}}\mathcal{L}_{\mathrm{pair}}
  (\vs_{\mathrm{student}},\vs_{\mathrm{teacher}})\\
  &+\lambda_{\mathrm{soft}}\mathcal{L}_{\mathrm{soft}}
  (\vs_{\mathrm{student}},\vs_{\mathrm{teacher}}),
  \end{aligned}
  \label{eq:speaker_distillation}
\end{equation}
where $\mathcal{L}_{\mathrm{cos}}$ aligns each student embedding with its
teacher target, $\mathcal{L}_{\mathrm{pair}}$ matches pairwise speaker
similarities within a minibatch, and $\mathcal{L}_{\mathrm{soft}}$ matches the
teacher-induced similarity distribution.  Only the student encoder is retained
for TTS, and we use $\vs\equiv\vs_{\mathrm{student}}$ as the speaker condition.
A learned projection inserts $\vs$ into the causal backbone, while
adaptive layer normalization conditions the diffusion head on the same
embedding.  Condition dropout provides the unconditional predictions used for
classifier-free guidance.

\subsection{Patch-based diffusion autoregression}
\label{sec:backbone}

Let $\vz_0,\vz_1,\ldots,\vz_{T-1}$ denote the target latent frames, and let
$K$ be the number of frames per patch.  After padding the sequence to a
multiple of $K$ when necessary, we form $N$ non-overlapping patches,
\begin{equation}
  \vP_i=\left(
  \vz_{iK},\vz_{iK+1},\ldots,\vz_{(i+1)K-1}
  \right),
  \qquad i\in\{0,\ldots,N-1\},
  \label{eq:latent_patch}
\end{equation}
where $N$ is the number of patches.  \cutetts{} uses $K=2$, giving
$\vP_i=(\vz_{2i},\vz_{2i+1})$.  The patch encoder $A$ is a bidirectional
Transformer applied independently within each patch.  It prepends a learned
summary token to the latent frames, allows the token to attend to the entire
patch, and uses its final state as the patch embedding $\va_i=A(\vP_i)$.  We
partition the reference latent sequence in the same way, denoting its patches
by $\vP^{\mathrm{ref}}_j$.  The same encoder yields the reference conditioning
sequence $\va^{\mathrm{ref}}=(A(\vP^{\mathrm{ref}}_0),\ldots,
A(\vP^{\mathrm{ref}}_{M-1}))$, where $M$ is the number of reference patches.
A causal Transformer summarizes the text, speaker condition, reference
sequence, and previously generated patch embeddings at step $i$:
\begin{equation}
  \vh_i
  =F_{\theta}\!\left(
  \vy,\vs,\va^{\mathrm{ref}},\va_{<i}
  \right).
  \label{eq:ar_context}
\end{equation}
The diffusion head then models the next patch conditioned on $\vh_i$, $\vs$,
and the preceding latent patch $\vP_{i-1}$.  This division of labor
follows the patch-based formulation of DiTAR~\citep{jia2025ditar}: the causal
backbone captures inter-patch structure, while bidirectional attention within
the diffusion head captures correlations inside a patch.

The diffusion head is trained with conditional flow
matching~\citep{lipman2023flowmatching}.
For a clean patch $\vP$ and Gaussian noise $\vxi$, we use the linear path
\begin{equation}
  \vx_t=(1-t)\vxi+t\vP, \qquad
  \vv^{\star}=\vP-\vxi,
  \label{eq:flow_path}
\end{equation}
and regress the conditional velocity
\begin{equation}
  \mathcal{L}_{\mathrm{FM}}
  =\mathbb{E}\left[
  \left\|\vv_{\theta}
  (\vx_t,t\mid\vh_i,\vs,\vP_{i-1})-\vv^{\star}\right\|_2^2
  \right].
  \label{eq:flow_loss}
\end{equation}
After sampling $\vP_i$, the patch encoder produces $\va_i$ for the next
autoregressive step.  A stop head terminates generation, and the VAE decoder
emits the corresponding waveform chunk.

\subsection{Guidance--step distillation}
\label{sec:distillation}

We use LM-level classifier-free guidance~\citep{ho2022cfg} to construct the
teacher trajectory.  For the conditional branch, let
$c_i=(\vh_i,\vs,\vP_{i-1})$ denote the patch condition, and let
$\vv_T(\vx,t\mid c_i)$ be the corresponding teacher velocity.  The
unconditional branch removes the text, reference, and speaker conditions while
retaining the generated patch history:
\begin{equation}
  \vh_i^{\mathrm{u}}
  =F_{\theta}\!\left(\va_{<i}\right),
  \qquad
  c_i^{\mathrm{u}}
  =\left(\vh_i^{\mathrm{u}},\vP_{i-1}\right).
  \label{eq:unconditional_context}
\end{equation}
The CFG-guided teacher velocity is then
\begin{equation}
  \vv_T^{\mathrm{cfg}}(\vx,t\mid c_i)
  =\vv_T(\vx,t\mid c_i)
  +w\left[
  \vv_T(\vx,t\mid c_i)-\vv_T(\vx,t\mid c_i^{\mathrm{u}})
  \right],
  \label{eq:cfg}
\end{equation}
where $w$ is the guidance scale.  Computing this velocity requires one
conditional and one unconditional forward pass at each solver step.

We distill the resulting teacher trajectory directly.  Starting from a noised
patch $\vx_t$ defined as in Equation~\ref{eq:flow_path}, let
$\vx_{t+\Delta t}^{\mathrm{teacher}}$ denote the endpoint obtained by
numerically integrating the CFG-guided teacher velocity over
$[t,t+\Delta t]$, where $\Delta t\in\mathcal{D}$ and $\mathcal{D}$ denotes the
supported interval lengths.  The student is conditioned on the starting time,
interval length, and guidance scale, and predicts the teacher's average
velocity over the interval,
\begin{equation}
  \bar{\vv}_T
  =\frac{\vx_{t+\Delta t}^{\mathrm{teacher}}-\vx_t}{\Delta t},
  \qquad
  \mathcal{L}_{\mathrm{GS}}
  =\mathbb{E}\left[
  \left\|\vv_S(\vx_t,t,\Delta t\mid c_i,w)-\bar{\vv}_T\right\|_2^2
  \right].
  \label{eq:joint_distill}
\end{equation}
At inference, a single student forward pass maps $\vx_t$ directly to
$\vx_{t+\Delta t}$,
\begin{equation}
  \vx_{t+\Delta t}=\vx_t+\Delta t\,
  \vv_S(\vx_t,t,\Delta t\mid c_i,w).
  \label{eq:student_update}
\end{equation}
The target in Equation~\ref{eq:joint_distill} is already CFG-guided, so the
student does not require a separate unconditional branch.  Because the interval
is an input, the same diffusion head can be evaluated with different step
budgets.  In
this sense, guidance removal and step reduction are learned jointly rather than
as two sequential distillation stages.

\section{Experiments}
\label{sec:experiments}

\subsection{Experimental Setup}

\paragraph{Implementation details.}

The TTS backbone contains approximately 230M parameters and comprises a
bidirectional patch encoder, a causal language-model backbone, and a bidirectional
diffusion head.  The audio VAE encodes 24-kHz speech into 64-dimensional latents at 12.5~Hz.
We group two frames per patch, resulting in an LM token rate of 6.25 tokens per second.

For each generated patch, the base system uses 10 diffusion-head function
evaluations (NFEs) on each of the two CFG branches, whereas the student uses
4 NFEs on a single branch.  Further training and inference details are provided
in Appendix~\ref{app:training_inference_details}.

\paragraph{Training and evaluation datasets.}

\cutetts{}, including its audio VAE and speaker encoder, is trained on an
internal multilingual dataset containing approximately 550,000 hours of speech.
For objective evaluation, we use LibriSpeech test-clean
\citep{panayotov2015librispeech} under the protocol adopted by
F5-TTS~\citep{chen2025f5tts}, together with the English and Chinese test sets
released with Seed-TTS~\citep{anastassiou2024seedtts}.  For subjective
evaluation and efficiency benchmarking, we sample 50 examples from LibriSpeech
test-clean and refer to this fixed set as the LibriSpeech test-clean subset.

\paragraph{Evaluation metrics.}

For objective evaluation, we compute Word Error Rate (WER) using
Faster-Whisper-large-v3 for LibriSpeech test-clean, Whisper-large-v3 for
Seed-TTS EN, and Paraformer-zh for Seed-TTS
ZH~\citep{radford2023whisper,gao2022paraformer}.  Speaker similarity (SIM) is
measured as the cosine similarity between embeddings from a fine-tuned
WavLM-Large model. For the ablation analyses, we additionally use
UTMOS~\citep{saeki2022utmos} as an automatic estimate of perceived speech
quality. For subjective evaluation, annotators compare paired
outputs from two systems along two dimensions: sound quality and naturalness.

\paragraph{Baselines.}

We compare \cutetts{} against a broad range of recent publicly available TTS
systems, including discrete-autoregressive systems such as the MOSS-TTS
family~\citep{gong2026mosstts}, Qwen3-TTS~\citep{hu2026qwen3tts}, and
FireRedTTS-2~\citep{xie2025fireredtts2}; continuous-non-autoregressive systems
such as F5-TTS~\citep{chen2025f5tts} and ZipVoice~\citep{zhu2025zipvoice};
hybrid systems such as IndexTTS2~\citep{zhou2025indextts2} and
CosyVoice~3~\citep{du2025cosyvoice3}; and continuous-autoregressive systems such
as VoxCPM2~\citep{zhou2026voxcpm2}, the VibeVoice family
\citep{peng2025vibevoice,microsoft2025vibevoicerealtime},
DiTAR~\citep{jia2025ditar}, and Pocket TTS~\citep{kyutai2026pockettts}.  All
baseline results were obtained using official implementations with default
settings, or as reported in their original papers.

\subsection{Experimental Results}
\label{sec:quality_results}

\paragraph{Objective evaluation.}

Table~\ref{tab:main_quality} compares \cutetts{} with publicly available systems
across three zero-shot voice-cloning benchmarks.  \cutetts{} achieves performance
comparable to leading systems with only 0.2B parameters, while maintaining low
recognition error and strong speaker similarity across all three test sets.  The
distilled student closely matches the base model across the evaluated metrics
while using a substantially smaller inference budget.
\begin{table*}[t]
\centering
\caption{Objective evaluation results of zero-shot voice-cloning performance.
``Disc.'' and ``Cont.'' denote discrete and continuous speech representations;
``AR'' and ``NAR'' denote autoregressive and non-autoregressive generation.
$\uparrow$ and $\downarrow$ indicate that higher and lower values are better,
respectively.  A dash denotes an unavailable result.}
\label{tab:main_quality}
\vspace{0.5em}
\setlength{\tabcolsep}{3.0pt}
\small
\begin{NoHyper}
\resizebox{\textwidth}{!}{%
\begin{tabular}{l!{\hspace{3pt}}l!{\hspace{3pt}}c!{\hspace{4pt}}cc cc cc}
\toprule
\multirow{2}{*}{Type} & \multirow{2}{*}{Model} & \multirow{2}{*}{Params.}
& \multicolumn{2}{c}{LibriSpeech test-clean}
& \multicolumn{2}{c}{Seed-TTS EN}
& \multicolumn{2}{c}{Seed-TTS ZH} \\
\cmidrule(lr){4-5} \cmidrule(lr){6-7} \cmidrule(lr){8-9}
& & & WER (\%)$\downarrow$ & SIM$\uparrow$ & WER (\%)$\downarrow$ & SIM$\uparrow$ & WER (\%)$\downarrow$ & SIM$\uparrow$ \\
\midrule
\multirow{4}{*}{\discar}
 & MOSS-TTS & 8B & 1.98 & 67.7 & 1.84 & 70.9 & 1.37 & 77.0 \\
 & Qwen3-TTS & 1.7B & 2.35 & 70.3 & 1.66 & 71.4 & \textbf{0.91} & 77.0 \\
 & FireRedTTS-2 & 1.5B & 4.32 & 64.2 & 1.95 & 66.5 & 1.14 & 73.6 \\
 & MOSS-TTS-Nano & 0.1B & 4.10 & 48.4 & 4.62 & 49.9 & 3.13 & 64.3 \\
\midrule
\multirow{2}{*}{\contnar}
 & F5-TTS & 0.3B & 2.42 & 66.0 & 1.83 & 67.0 & 1.56 & 76.0 \\
 & ZipVoice & 0.1B & 2.05 & 67.4 & 1.70 & 69.7 & 1.40 & 75.1 \\
\midrule
\multirow{2}{*}{\hybridar}
 & IndexTTS2 & 1.5B & 2.47 & 70.0 & 2.22 & 70.6 & 1.02 & 76.5 \\
 & CosyVoice~3 & 0.5B & 1.99 & 69.7 & 2.02 & 71.8 & 1.16 & 78.0 \\
\midrule
\multirow{5}{*}{\contar}
 & VoxCPM2 & 2B & 3.01 & 74.0 & 1.84 & 75.3 & 0.97 & \textbf{79.5} \\
 & VibeVoice & 1.5B & -- & -- & 3.04 & 68.9 & 1.16 & 74.4 \\
 & DiTAR & 0.6B & 2.39 & 67.0 & 1.69 & 73.5 & 1.02 & 75.3 \\
 & VibeVoice-Realtime & 0.5B & 2.00 & 69.5 & 2.05 & 63.3 & -- & -- \\
 & Pocket TTS & 0.1B & \textbf{1.59} & 49.1 & \textbf{1.63} & 50.7 & -- & -- \\
\midrule
\midrule
\multirow{2}{*}{\contar}
 & \textbf{\modelbase{}} & 0.2B & 2.16 & \textbf{78.9} & 2.04 & \textbf{76.5} & 1.41 & 77.8 \\
 & \textbf{\modeldistill{}} & 0.2B & 2.41 & 76.8 & 2.03 & 74.2 & 1.47 & 75.6 \\
\bottomrule
\end{tabular}}
\end{NoHyper}
\end{table*}

\paragraph{Subjective evaluation.}

On the LibriSpeech test-clean subset, we compare \cutetts{} and
\modeldistill{} with four similarly sized TTS systems: MOSS-TTS-Nano, F5-TTS,
ZipVoice, and Pocket TTS.  Annotators evaluate paired outputs independently in
terms of sound quality and naturalness under a blinded protocol.  Further
details are provided in Appendix~\ref{app:subjective_evaluation_details}.

\begin{figure*}[t]
  \centering
  \includegraphics[width=0.94\textwidth]{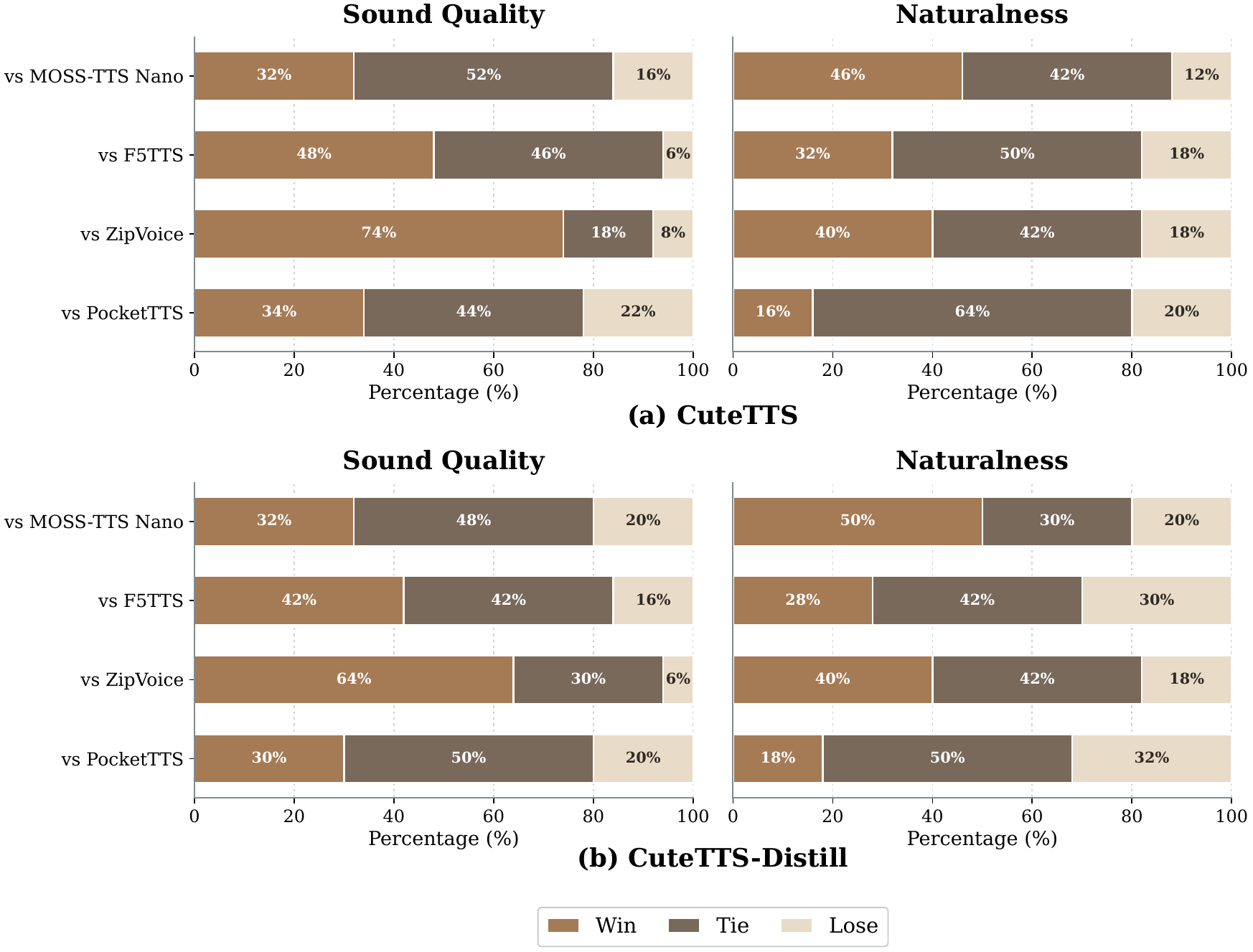}
  \caption{Subjective evaluation on the LibriSpeech test-clean
  subset.  Results are reported as win/tie/loss percentages from the
  perspective of each \cutetts{} variant.}
  \label{fig:subjective_evaluation}
\end{figure*}

Figure~\ref{fig:subjective_evaluation} reports win, tie, and loss percentages
from the perspective of each \cutetts{} variant.  Both variants are preferred
over all four baselines in terms of sound quality.  The base model also receives
more naturalness wins than losses against MOSS-TTS-Nano, F5-TTS, and ZipVoice,
while its comparison with Pocket TTS is dominated by ties.  The distilled model
yields results closely matching those of the base model across both evaluation
dimensions.

\subsection{Efficiency}
\label{sec:efficiency_results}

We define first-audio latency as the elapsed time from raw target text and a
reference-audio path to the first PCM chunk on the CPU, and real-time factor
(RTF) as total synthesis time divided by generated-audio duration. We report the
mean, P50, and P95 first-audio latency, where P50 is the median and P95 is the
latency threshold below which 95\% of requests fall, characterizing tail
latency. All efficiency measurements are conducted on an NVIDIA RTX~4090.
Further benchmark details are provided in
Appendix~\ref{app:efficiency_benchmark}.

Table~\ref{tab:streaming_efficiency} compares end-to-end inference efficiency
across representative system families.  For \cutetts{}, guidance--step
distillation reduces mean first-audio latency by 23.3\% and mean RTF by 40.8\%
under this paired protocol. It also lowers P50 latency from 48.7 to 37.1~ms and
P95 latency from 52.3 to 40.8~ms, showing consistent gains in both typical and
tail latency.

\begin{table*}[t]
\centering
\caption{End-to-end efficiency across representative TTS systems, measured by
first-audio latency and real-time factor (RTF).}
\label{tab:streaming_efficiency}
\vspace{0.5em}
\setlength{\tabcolsep}{3.5pt}
\small
\resizebox{0.92\textwidth}{!}{%
\begin{tabular}{l!{\hspace{3pt}}l!{\hspace{3pt}}c!{\hspace{4pt}}rrr!{\hspace{8pt}}c}
\toprule
\multirow{2}{*}{Type} & \multirow{2}{*}{Model} & \multirow{2}{*}{Params.}
& \multicolumn{3}{c}{Latency (ms)$\downarrow$} & \multirow{2}{*}{RTF$\downarrow$} \\
\cmidrule(lr){4-6}
& & & Mean & P50 & P95 & \\
\midrule
\multirow{4}{*}{\discar}
 & MOSS-TTS (vLLM-Omni) & 8B & -- & -- & -- & 0.498 \\
 & Qwen3-TTS (vLLM-Omni) & 1.7B & 108.8 & 108.4 & 112.2 & 0.190 \\
 & FireRedTTS-2 & 1.5B & 316.0 & 314.7 & 326.6 & 1.113 \\
 & MOSS-TTS-Nano & 0.1B & 123.8 & 122.7 & 142.8 & 0.778 \\
\midrule
\multirow{2}{*}{\contnar}
 & F5-TTS & 0.3B & 628.2 & 622.2 & 667.5 & 0.122 \\
 & ZipVoice & 0.1B & 200.5 & 191.3 & 250.0 & \textbf{0.030} \\
\midrule
\multirow{2}{*}{\hybridar}
 & IndexTTS2 & 1.5B & 3826.1 & 3694.8 & 5490.0 & 0.676 \\
 & CosyVoice~3 & 0.5B & 2370.1 & 2263.7 & 3345.4 & 0.453 \\
\midrule
\multirow{3}{*}{\contar}
 & VoxCPM2 & 2B & 66.6 & 63.0 & 82.2 & 0.404 \\
 & VibeVoice-Realtime & 0.5B & 62.9 & 62.6 & 64.8 & 0.365 \\
 & Pocket TTS & 0.1B & 51.0 & 50.7 & 52.9 & 0.125 \\
\midrule
\midrule
\multirow{2}{*}{\contar}
 & \textbf{\modelbase{}} & 0.2B & 49.0 & 48.7 & 52.3 & 0.184 \\
 & \textbf{\modeldistill{}} & 0.2B & \textbf{37.6} & \textbf{37.1} & \textbf{40.8} & 0.109 \\
\bottomrule
\end{tabular}}
\end{table*}

\section{Ablation Studies and Analysis}
\label{sec:ablations}

All ablation experiments use an approximately 50,000-hour subset of the full
training corpus. Detailed experimental configurations are provided in
Appendix~\ref{app:ablation_experiments}.

\subsection{Temporal Granularity in Continuous Autoregressive Modeling}

The VAE frame rate $r$ and patch size $p$ jointly determine the temporal
resolution of the speech representation and the granularity of autoregressive
prediction. Given a VAE frame rate of $r$ frames per second and $p$ frames per
patch, the LM token rate $q$, defined as the number of autoregressive steps taken
by the causal backbone per second, is
\begin{equation}
  q = \frac{r}{p}
\end{equation}
A higher $r$ provides finer temporal detail but increases the latent sequence
length, whereas a larger $p$ lowers the LM token rate but requires the diffusion
head to generate a larger continuous patch at each step. Different combinations
of $r$ and $p$ can yield the same LM token rate while differing in representation
resolution and patch-generation difficulty. We therefore first examine two controlled
settings: varying $r$ with $p$ fixed, and varying $r$ and $p$ while holding $q$
constant. We then complement these comparisons with module-wise depth scaling
to examine how different temporal granularities interact with the capacities
of the causal backbone and diffusion head.

\subsubsection{Representation Rate and Reconstruction Quality}
\label{sec:representation_results}

\begin{figure}[t]
  \centering
  \includegraphics[width=0.72\linewidth]{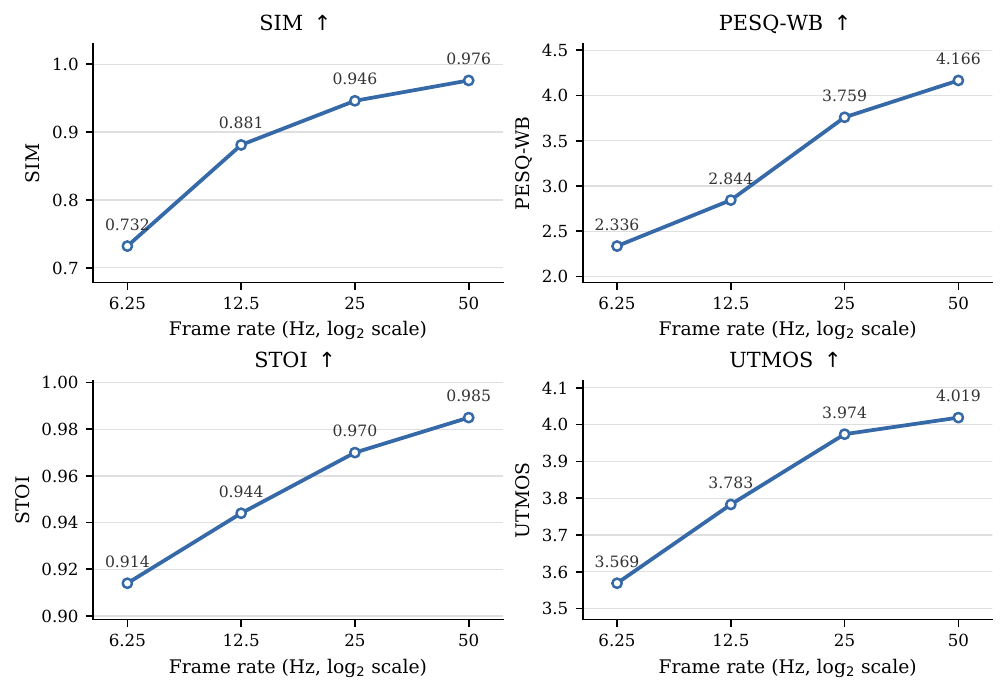}
  \caption{VAE reconstruction quality at different frame rates on
    LibriSpeech test-clean. All models use the same training data and training
  configuration.}
  \label{fig:vae_frame_rate}
\end{figure}

To study the effect of the VAE frame rate $r$, we train four VAEs with
$r \in \{6.25, 12.5, 25, 50\}$ Hz using the same training data and training
configuration. The models differ only in the downsampling factor of the
encoder's final stage and the corresponding upsampling factor of the decoder's
final stage. Their reconstruction performance on LibriSpeech test-clean is
visualized in Figure~\ref{fig:vae_frame_rate}. Reconstruction quality improves
consistently as $r$ increases across all four metrics. The improvement is
steepest from 6.25 Hz to 12.5 Hz and gradually diminishes at higher frame
rates.

\subsubsection{Joint Effects of VAE Frame Rate and Patch Size}

We denote a TTS model with VAE frame rate $r$ and patch size $p$ as
$\mathrm{TTS}_{r,p}$, where $r$ is measured in Hz and $p$ is the number of VAE
frames in each patch. Using a training configuration shared across all
variants, we train the configurations reported below and evaluate them on
LibriSpeech test-clean.

\textbf{Fixed patch size.}
We first fix $p=1$, for which the LM token rate grows directly with the VAE frame
rate. Table~\ref{tab:fixed_patch_size} shows that increasing $r$ initially
improves SIM and UTMOS, consistent with the stronger reconstruction capability
of the higher-rate VAEs. However, WER rises rapidly because the causal backbone
must make proportionally more autoregressive decisions. At 50 Hz, WER increases
sharply and both SIM and UTMOS decline relative to the intermediate-rate
models, suggesting that the resulting temporal granularity exceeds the
modeling capacity of the fixed backbone.

\begin{table}[t]
  \centering
  \caption{TTS performance at a fixed patch size of $p=1$ on LibriSpeech
  test-clean.}
  \label{tab:fixed_patch_size}
  \vspace{0.5em}
  \normalsize
  \setlength{\tabcolsep}{8pt}
  \begin{tabular}{lrrr}
    \toprule
    Model & WER (\%)$\downarrow$ & SIM$\uparrow$ & UTMOS$\uparrow$ \\
    \midrule
    \textbf{$\mathrm{TTS}_{6.25,1}$} & \textbf{2.53} & 62.5 & 3.733 \\
    \textbf{$\mathrm{TTS}_{12.5,1}$} & 4.67 & \textbf{71.0} & \textbf{3.883} \\
    \textbf{$\mathrm{TTS}_{25,1}$} & 10.59 & \textbf{71.0} & 3.833 \\
    \textbf{$\mathrm{TTS}_{50,1}$} & 70.33 & 67.9 & 3.346 \\
    \bottomrule
  \end{tabular}
\end{table}

\textbf{Fixed LM token rate.}
To separate autoregressive sequence length from patch-generation difficulty,
we next fix the LM token rate at $q=r/p=6.25$ tokens per second. The four configurations in
Table~\ref{tab:fixed_token_rate} require the same number of autoregressive
decisions but use progressively higher VAE frame rates and larger patches.
$\mathrm{TTS}_{12.5,2}$ improves SIM and UTMOS substantially over
$\mathrm{TTS}_{6.25,1}$ with only a small change in WER, showing that a
moderate patch size can exploit the stronger reconstruction capability of a
higher-rate VAE without materially degrading intelligibility. With still
larger patches, SIM continues to improve, but WER increases and UTMOS declines.
This divergence shows that LM token rate alone does not characterize synthesis
difficulty: larger patches reduce the autoregressive sequence length but alter
the relative demands on inter-patch autoregressive modeling and within-patch
continuous generation. We next examine this interaction through module-wise
capacity scaling.

\begin{table}[t]
  \centering
  \caption{TTS performance at a fixed LM token rate of $q=6.25$ tokens per second
  on LibriSpeech test-clean.}
  \label{tab:fixed_token_rate}
  \vspace{0.5em}
  \normalsize
  \setlength{\tabcolsep}{8pt}
  \begin{tabular}{lrrr}
    \toprule
    Model & WER (\%)$\downarrow$ & SIM$\uparrow$ & UTMOS$\uparrow$ \\
    \midrule
    \textbf{$\mathrm{TTS}_{6.25,1}$} & \textbf{2.53} & 62.5 & 3.733 \\
    \textbf{$\mathrm{TTS}_{12.5,2}$} & 2.70 & 71.3 & \textbf{3.878} \\
    \textbf{$\mathrm{TTS}_{25,4}$} & 3.92 & 72.5 & 3.749 \\
    \textbf{$\mathrm{TTS}_{50,8}$} & 7.04 & \textbf{74.1} & 3.649 \\
    \bottomrule
  \end{tabular}
\end{table}

\subsubsection{Module-wise Capacity Sensitivity}
\label{sec:capacity_sensitivity}

The preceding experiments show that temporal granularity affects both
autoregressive sequence modeling and within-patch generation. To further
understand how these two modeling demands interact with model capacity, we
consider two configurations using the same 50-Hz speech representation but
substantially different patch sizes, $\mathrm{TTS}_{50,1}$ and
$\mathrm{TTS}_{50,8}$. They respectively represent a fine-grained regime with
a small patch and a long autoregressive sequence, and a coarse-grained regime
with a large patch and a much shorter autoregressive sequence. Starting from
the base architecture, we independently double the depth of either the causal
backbone or the diffusion head while keeping the other component unchanged.
All other training and inference configurations are shared across variants.
Detailed model configurations are provided in
Appendix~\ref{app:ablation_experiments}. Since doubling the depth of the two
modules results in different increases in parameter count, we interpret this
experiment as a module-wise sensitivity analysis rather than a
parameter-matched capacity comparison.

\begin{table}[t]
  \centering
  \caption{Module-wise depth scaling under two temporal granularities with a
    fixed 50-Hz speech representation. ``Causal backbone $\times2$'' and
    ``Diffusion head $\times2$'' independently double the depth of the
  corresponding module.}
  \label{tab:capacity_scaling}
  \vspace{0.5em}
  \normalsize
  \setlength{\tabcolsep}{7pt}
  \begin{tabular}{lrrr}
    \toprule
    Model & WER (\%)$\downarrow$ & SIM$\uparrow$ & UTMOS$\uparrow$ \\
    \midrule\midrule
    $\mathrm{TTS}_{50,1}$ & 70.33 & 67.9 & 3.346 \\
    \midrule
    \quad Causal backbone $\times2$ & \textbf{60.03} & 72.1 & \textbf{3.434} \\
    \quad Diffusion head $\times2$ & 90.84 & \textbf{72.4} & 3.419 \\
    \midrule\midrule
    $\mathrm{TTS}_{50,8}$ & 7.04 & 74.1 & 3.649 \\
    \midrule
    \quad Causal backbone $\times2$ & \textbf{4.91} & 74.4 & 3.700 \\
    \quad Diffusion head $\times2$ & 5.85 & \textbf{76.8} & \textbf{3.924} \\
    \bottomrule
  \end{tabular}
\end{table}

Table~\ref{tab:capacity_scaling} shows that, for the fine-grained
$\mathrm{TTS}_{50,1}$ configuration, the causal backbone operates at 50
autoregressive steps per second. Doubling its depth reduces WER
from 70.33\% to 60.03\%, while also improving SIM and UTMOS. In contrast,
enlarging the diffusion head improves speaker similarity but does not alleviate
the severe intelligibility degradation, with WER further increasing to
90.84\%. This suggests that, when autoregressive prediction is extremely
fine-grained, the long sequence places substantial modeling pressure on the
causal backbone, and additional local-generation capacity alone cannot
compensate for this difficulty.

A different pattern emerges for $\mathrm{TTS}_{50,8}$, where patching reduces
the LM token rate from 50 to 6.25 tokens per second. Increasing causal-backbone
depth primarily improves intelligibility, reducing WER from 7.04\% to 4.91\%,
while producing only marginal gains in SIM and UTMOS. In contrast, doubling the
diffusion-head depth produces substantially larger improvements in speaker
fidelity and predicted perceptual quality: SIM increases from 74.1 to 76.8 and
UTMOS from 3.649 to 3.924, while WER also improves to 5.85\%. This indicates
that, once the autoregressive sequence is substantially shortened and each
autoregressive step is responsible for a larger continuous patch, additional
diffusion-head capacity becomes increasingly useful for local acoustic
generation.

Taken together, the two configurations exhibit complementary module-wise
capacity sensitivities. Increasing causal-backbone capacity primarily benefits
intelligibility, consistent with its role in modeling inter-patch
dependencies, whereas increasing diffusion-head capacity more strongly
benefits speaker fidelity and perceptual quality, consistent with its role in
within-patch continuous generation. Temporal granularity further modulates the
relative importance of these capacities: causal-backbone capacity is
particularly important for fine-grained autoregressive prediction, while
diffusion-head capacity becomes more valuable under large-patch generation.
Temporal granularity therefore controls more than the autoregressive token
rate: it also changes how modeling capacity is utilized across inter-patch
autoregression and within-patch continuous generation.

\subsection{Effect of Semantic Alignment}

To assess the effect of semantic alignment, we train an additional VAE using
the same configuration as the 12.5-Hz VAE in Section~\ref{sec:representation_results},
but without the semantic-alignment objective. Table~\ref{tab:semantic_alignment_reconstruction}
shows that adding semantic alignment incurs virtually no degradation in
reconstruction quality.

\begin{table}[t]
  \centering
  \caption{Effect of semantic alignment on VAE reconstruction at a frame rate of
  12.5 Hz.}
  \label{tab:semantic_alignment_reconstruction}
  \vspace{0.5em}
  \small
  \setlength{\tabcolsep}{5pt}
  \resizebox{0.82\linewidth}{!}{%
    \begin{tabular}{lrrrrr}
      \toprule
      Model & Rate (Hz) & SIM$\uparrow$ & PESQ-WB$\uparrow$ & STOI$\uparrow$ & UTMOS$\uparrow$ \\
      \midrule
      w/ semantic & 12.5 & 0.881 & 2.844 & 0.944 & 3.783 \\
      w/o semantic & 12.5 & 0.881 & 2.885 & 0.945 & 3.786 \\
      \bottomrule
  \end{tabular}}
\end{table}

\begin{figure}[t]
  \centering
  \includegraphics[width=0.96\linewidth]{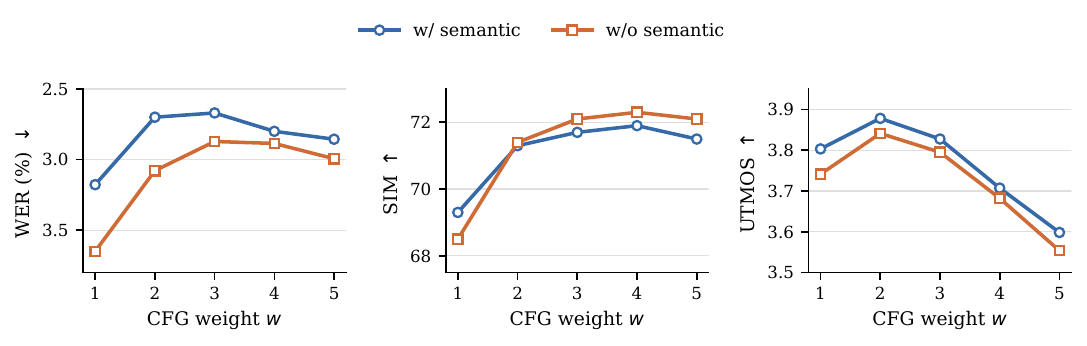}
  \caption{Effect of semantic alignment on TTS evaluation metrics across CFG
    weights on LibriSpeech test-clean. The WER axis is reversed so that better
  performance is shown higher.}
  \label{fig:semantic_alignment_cfg}
\end{figure}

We then train a TTS model on the representation without semantic alignment
using the same configuration as $\mathrm{TTS}_{12.5,2}$.
Figure~\ref{fig:semantic_alignment_cfg} compares the two models across CFG
weights. The semantically aligned representation substantially reduces WER
across guidance strengths while retaining comparable SIM and UTMOS. These
results indicate that semantic alignment makes the continuous representation
easier for the causal backbone to model, improving intelligibility without
compromising speaker similarity or predicted synthesis quality.

\subsection{Effect of Explicit Speaker Conditioning}

The default $\mathrm{TTS}_{12.5,2}$ uses the explicit speaker embedding in
both the LM prefix and the diffusion head, with adaptive layer normalization
(AdaLN) providing the diffusion-head condition. To assess the contribution of
explicit speaker conditioning, we train two variants using the same training
configuration: $\mathrm{TTS}_{12.5,2}$ (no spk), which removes the explicit
speaker embedding from both components, and $\mathrm{TTS}_{12.5,2}$
(diff. head spk), which provides it only to the diffusion head. To examine the
injection mechanism, we also train $\mathrm{TTS}_{12.5,2}$ (spk as global
token), which retains LM-prefix speaker conditioning but injects the speaker
embedding into the diffusion head as a global token rather than through AdaLN.

We compare all variants on LibriSpeech test-clean with CFG weight $w=2$.
Because our speaker encoder is distilled from WavLM and the primary SIM
evaluator is also WavLM-based, proximity between their representation spaces
could influence the measured similarity. We therefore additionally report SIM
using WeSpeaker~\citep{wang2023wespeaker} as an independent speaker-embedding
model.

\begin{table}[t]
  \centering
  \caption{Effect of explicit speaker conditioning on $\mathrm{TTS}_{12.5,2}$
  performance at CFG weight $w=2$ on LibriSpeech test-clean.}
  \label{tab:speaker_conditioning_ablation}
  \vspace{0.5em}
  \normalsize
  \setlength{\tabcolsep}{5pt}
  \begin{tabular}{@{}lrrr@{}}
    \toprule
    Model & WER (\%)$\downarrow$ & SIM$\uparrow$ & SIM (WeSpeaker)$\uparrow$ \\
    \midrule
    \textbf{$\mathrm{TTS}_{12.5,2}$}
    & 2.70 & \textbf{71.3} & \textbf{82.5} \\
    \midrule
    \quad\textit{no spk}
    & 2.76 & 53.4 & 74.5 \\
    \quad\textit{diff. head spk}
    & \textbf{2.52} & 71.2 & \textbf{82.5} \\
    \quad\textit{spk as global token}
    & 2.76 & 71.0 & 82.4 \\
    \bottomrule
  \end{tabular}
\end{table}

As shown in Table~\ref{tab:speaker_conditioning_ablation}, removing the explicit
speaker embedding substantially degrades similarity under both speaker
evaluators, confirming its importance for zero-shot voice cloning. Conditioning
the diffusion head alone recovers most of this loss, and using the speaker
embedding as a global token instead of AdaLN provides no consistent benefit.
The default $\mathrm{TTS}_{12.5,2}$ achieves
the highest WavLM-based SIM, ties the best WeSpeaker SIM, and remains
competitive in WER. It therefore provides the strongest overall balance and
is retained as our default conditioning design.

\subsection{Effect of Guidance--Step Distillation}

\begin{figure}[t]
  \centering
  \includegraphics[width=0.98\linewidth]{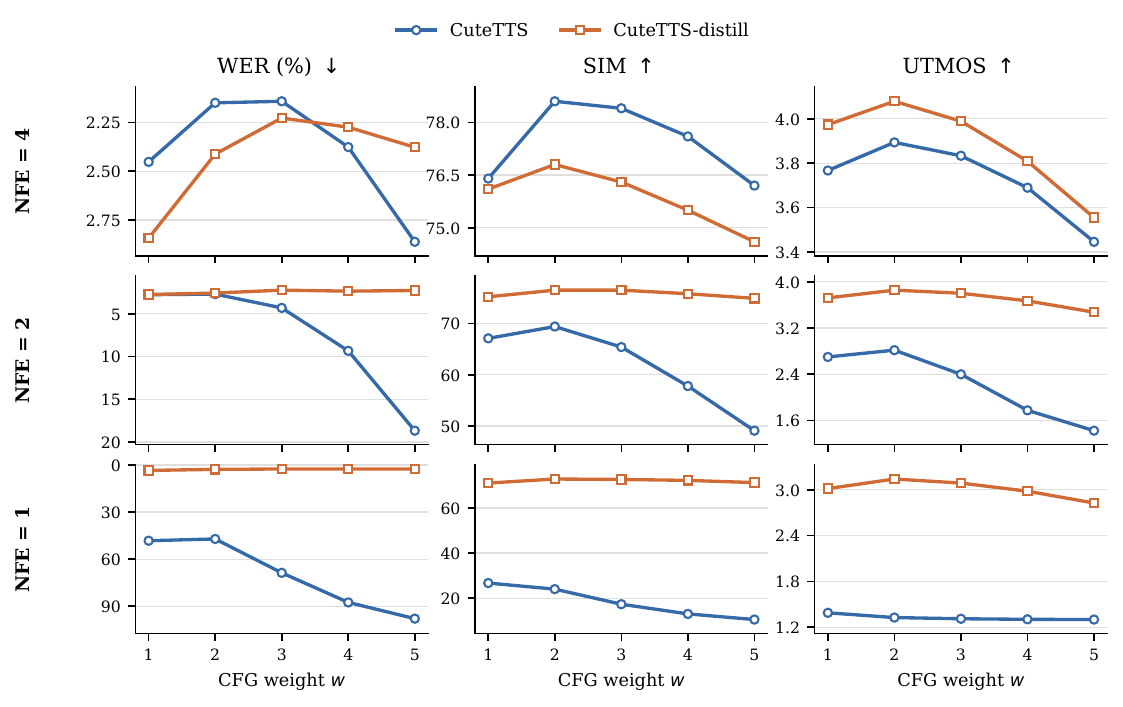}
  \caption{Effect of guidance--step distillation on TTS performance across CFG
    weights and diffusion-head NFE budgets on LibriSpeech test-clean. WER axes
  are reversed so that better performance is shown higher.}
  \label{fig:guidance_step_distillation}
\end{figure}

To assess guidance--step distillation across inference budgets, we compare
\modelbase{} and \modeldistill{} using 4, 2, and 1 diffusion-head NFEs and
sweep the CFG weight from 1 to 5 on LibriSpeech test-clean. The base model uses
standard two-branch LM-CFG with a Sway coefficient of $-0.8$, whereas the
distilled model uses its single-branch distilled guidance and a uniform time
grid. Figure~\ref{fig:guidance_step_distillation} reports WER, SIM, and UTMOS
for all settings.

At four NFEs, both models maintain strong performance over moderate guidance
weights. As the NFE budget decreases, the base model degrades sharply, while
the distilled model retains low WER and high speaker similarity at both two
and one NFE. The distilled model always achieves higher UTMOS than the base
model across the tested NFE budgets and CFG weights. These results show that guidance--step
distillation substantially improves robustness under highly constrained
sampling budgets.

\section{Conclusion}
\label{sec:conclusion}

We presented \cutetts, a compact streaming TTS system that combines patch-level autoregression over continuous speech latents with a bidirectional flow-matching head. Its semantically aligned causal $\sigma$-VAE preserves acoustic detail at a low rate, while explicit speaker conditioning supplies a direct identity cue for zero-shot voice cloning. Across LibriSpeech and Seed-TTS benchmarks, \cutetts{} achieves competitive intelligibility and speaker similarity, with paired subjective evaluations showing a consistent sound-quality preference over similarly sized baselines.

Our controlled studies reveal that representation rate and generation granularity must be considered jointly. Higher-rate latents improve reconstruction but increase the prediction burden, whereas moderate patching retains much of their acoustic benefit without inflating the language-model token rate. Module-wise scaling further reveals complementary capacity sensitivities between the two generative components. Increasing causal-backbone capacity primarily improves intelligibility, whereas increasing diffusion-head capacity more strongly benefits speaker similarity and perceptual quality. Fine-grained autoregressive prediction is particularly sensitive to backbone capacity, while additional diffusion-head capacity becomes increasingly useful under large-patch generation. Semantic alignment improves intelligibility, and explicit speaker conditioning preserves identity. Finally, guidance--step distillation directly addresses the inference cost of two-branch CFG and iterative patch generation. It compresses two-branch CFG and multiple integration steps into a single head that remains effective at small NFE budgets. Under the paired efficiency protocol, the distilled model reduces mean first-audio latency by $23.3\%$ and mean real-time factor by $40.8\%$ while maintaining synthesis quality close to the base model. These findings provide a coherent design principle for balancing representation fidelity, autoregressive granularity, module-wise capacity, and sampling cost in streaming continuous-autoregressive TTS.

\bibliography{references}
\bibliographystyle{colm2024_conference}

\clearpage
\appendix
\section*{Appendix}
\section{Training and Inference Details}
\label{app:training_inference_details}

\subsection{Model Configuration}

Tables~\ref{tab:audio_vae_configuration} and~\ref{tab:tts_configuration}
summarize the model configurations used in our experiments.
The audio VAE accepts and reconstructs 24-kHz waveforms. It uses a causal
convolutional architecture and a $\sigma$-VAE posterior with its standard
deviation fixed to 0.15. \modelbase{} and \modeldistill{} share the same
overall architecture, while the latter augments the diffusion head with
guidance-strength and step-size embeddings. Both models group two latent frames
per patch, resulting in an LM token rate of 6.25 tokens per second.

\subsection{Training Configuration}

\textbf{Audio VAE.}
Training is conducted in FP32 for 1M steps with an effective batch size of 256.
Training samples are randomly cropped into 2.5-s segments. To improve
reconstruction from bandwidth-degraded speech, we downsample 5\% of the
training examples to 16~kHz and another 5\% to 8~kHz, resample them back to
24~kHz, and retain the original 24-kHz signals as reconstruction targets. The
loss weights in Equation~\ref{eq:vae_objective} are $\lambda_{\mathrm{mel}}=15$,
$\lambda_{\mathrm{adv}}=1$, $\lambda_{\mathrm{feat}}=2$,
$\lambda_{\mathrm{KL}}=0.1$, and $\lambda_{\mathrm{sem}}=1$. We use AdamW with
a learning rate of $2\times10^{-4}$, $(\beta_1,\beta_2)=(0.8,0.99)$, and
weight decay of 0.01, followed by exponential decay with $\gamma=0.999996$.

\textbf{\modelbase{}.}
We train the TTS backbone end-to-end for 1M steps with a global batch size of
up to 81,920 packed tokens. For flow-matching training, each target patch is
repeated four times; each copy uses independently sampled Gaussian noise and
$t=\operatorname{sigmoid}(u)$, where $u\sim\mathcal{N}(0,1)$. Conditional inputs
are dropped with probability 0.1 to enable CFG at inference. We use AdamW with
a peak learning rate of $5\times10^{-4}$,
$(\beta_1,\beta_2)=(0.9,0.95)$, and weight decay of 0.01. The learning rate
follows a cosine schedule with 5K warmup steps.

\textbf{\modeldistill{}.}
We update only the diffusion head for 100K steps with a global batch size of up
to 65,536 packed tokens. Each target patch is repeated twice for distillation.
Teacher targets are computed on an eight-step base grid. At 0, 20K, and 50K
training steps, the sampling probabilities over four-, two-, and one-step
targets are $(1,0,0)$, $(0.5,0.5,0)$, and $(0.2,0.3,0.5)$, respectively; the
interval start is sampled uniformly from the selected discrete grid. We use
AdamW with a peak learning rate of $1\times10^{-5}$,
$(\beta_1,\beta_2)=(0.9,0.95)$, weight decay of 0.01, and a cosine schedule with
1K warmup steps. The resulting checkpoint supports all three inference budgets.

\begin{table}[h]
  \centering
  \caption{Audio VAE configuration.}
  \label{tab:audio_vae_configuration}
  \vspace{0.5em}
  \normalsize
  \setlength{\tabcolsep}{8pt}
  \renewcommand{\arraystretch}{1.12}
  \begin{tabular}{lr}
    \toprule
    Hyperparameter & Value \\
    \midrule
    Frame rate (Hz) & 12.5 \\
    Latent dimension & 64 \\
    Encoder initial channels & 128 \\
    Encoder strides & 3/5/8/16 \\
    Decoder initial channels & 1,536 \\
    Decoder strides & 16/8/5/3 \\
    Parameters (M) & 127.9 \\
    \bottomrule
  \end{tabular}
\end{table}

\begin{table*}[t]
  \centering
  \caption{Model configurations. Parameter counts are rounded.}
  \label{tab:tts_configuration}
  \vspace{0.5em}
  \small
  \setlength{\tabcolsep}{6pt}
  \begin{tabular}{llcc}
    \toprule
    Component & Hyperparameter & \textbf{\modelbase{}} & \textbf{\modeldistill{}} \\
    \midrule
    \multirow{5}{*}{Patch encoder}
    & Layers & 2 & 2 \\
    & Hidden dimension & 1,024 & 1,024 \\
    & Attention/KV heads & 16/2 & 16/2 \\
    & FFN dimension & 4,096 & 4,096 \\
    & Parameters (M) & 31.0 & 31.0 \\
    \midrule
    \multirow{5}{*}{Causal backbone}
    & Layers & 7 & 7 \\
    & Hidden dimension & 1,024 & 1,024 \\
    & Attention/KV heads & 16/8 & 16/8 \\
    & FFN dimension & 3,072 & 3,072 \\
    & Parameters (M) & 126.9 & 126.9 \\
    \midrule
    \multirow{7}{*}{Diffusion head}
    & Layers & 4 & 4 \\
    & Hidden dimension & 1,024 & 1,024 \\
    & Attention/KV heads & 16/2 & 16/2 \\
    & FFN dimension & 4,096 & 4,096 \\
    & Guidance-strength embedding & -- & $\checkmark$ \\
    & Step-size embedding & -- & $\checkmark$ \\
    & Parameters (M) & 70.5 & 73.6 \\
    \midrule
    Overall & Parameters (M) & 228.6 & 231.8 \\
    \bottomrule
  \end{tabular}
\end{table*}
\subsection{Inference Configuration}

\textbf{Voice-cloning CFG.}
The base model uses LM-level CFG with guidance weight $w=2$.
Table~\ref{tab:voice_clone_cfg_branches} summarizes the two branches. Both
branches retain the generated patch history. The unconditional branch removes
the target text and reference audio from the LM input and zeros the
diffusion-head speaker condition. CFG therefore jointly strengthens the
LM-side text and reference conditions and the diffusion-head speaker condition.

\begin{table}[t]
  \centering
  \caption{LM-CFG branches for zero-shot voice cloning with \modelbase{}.}
  \label{tab:voice_clone_cfg_branches}
  \vspace{0.5em}
  \small
  \setlength{\tabcolsep}{3pt}
  \begin{tabular}{@{}l>{\centering\arraybackslash}p{0.48\columnwidth}>{\centering\arraybackslash}p{0.23\columnwidth}@{}}
    \toprule
    Branch & LM input & Diffusion-head speaker \\
    \midrule
    Conditional
    & $\langle\text{reference audio}\rangle\mid\langle\text{text}\rangle$
    & $\vs$ \\
    Unconditional
    & Unconditional LM prefix
    & $\mathbf{0}$ \\
    \bottomrule
  \end{tabular}
\end{table}

\textbf{Sampling.}
For each generated patch, \modelbase{} uses 10 diffusion-head NFEs on each CFG
branch, with Euler integration over a Sway time grid with coefficient $-0.8$.
\modeldistill{} encodes $w=2$ directly as a diffusion-head condition and uses
four NFEs on a single uniform Euler grid, without an unconditional branch. The
same distilled checkpoint supports one-, two-, and four-step inference; unless
otherwise noted, the main experiments use four steps.

\subsection{Ablation Configuration}
\label{app:ablation_experiments}

\textbf{VAE.}
The VAE ablations accept and reconstruct 24-kHz waveforms and use the same
causal convolutional architecture, 64-dimensional $\sigma$-VAE posterior, and
fixed posterior standard deviation of 0.15 as the main audio VAE. As summarized
in Table~\ref{tab:vae_ablation_model_configuration}, the four frame-rate
variants differ only in the final encoder downsampling factor and the
corresponding decoder upsampling factor. All other model settings are shared.


\begin{table*}[t]
  \centering
  \caption{Model configurations for the VAE frame-rate ablation.}
  \label{tab:vae_ablation_model_configuration}
  \vspace{0.5em}
  \small
  \setlength{\tabcolsep}{7pt}
  \begin{tabular}{lcccc}
    \toprule
    Hyperparameter & 6.25 Hz & 12.5 Hz & 25 Hz & 50 Hz \\
    \midrule
    Latent dimension & 64 & 64 & 64 & 64 \\
    Encoder initial channels & 128 & 128 & 128 & 128 \\
    Encoder strides & 3/5/8/32 & 3/5/8/16 & 3/5/8/8 & 3/5/8/4 \\
    Decoder initial channels & 1,536 & 1,536 & 1,536 & 1,536 \\
    Decoder strides & 32/8/5/3 & 16/8/5/3 & 8/8/5/3 & 4/8/5/3 \\
    Parameters (M) & 232.7 & 127.9 & 75.5 & 49.2 \\
    \bottomrule
  \end{tabular}
\end{table*}

The four variants are trained for 500K steps in FP32 on randomly cropped
2.5-s segments with an effective batch size of 64. We optimize both the
generator and discriminator using AdamW with a learning rate of
$1\times10^{-4}$, $(\beta_1,\beta_2)=(0.8,0.99)$, and weight decay of 0.01,
followed by exponential decay with $\gamma=0.999996$. The weights for mel
reconstruction, adversarial feature matching, adversarial generation, KL
regularization, and semantic alignment are 15, 2, 1, 0.1, and 1, respectively.

\textbf{TTS.}
Unless otherwise specified, all TTS ablations use the model configuration in
Table~\ref{tab:tts_configuration}. The layer counts and hidden dimensions are
held fixed across the Rate--Patch, semantic-alignment, and speaker-conditioning
experiments, while the patch-encoder input and diffusion-head output projections
follow the patch size $p$. For the module-wise capacity study in
Section~\ref{sec:capacity_sensitivity}, we additionally construct variants by
independently doubling either the causal-backbone depth from 7 to 14 layers or
the diffusion-head depth from 4 to 8 layers while leaving the remaining
architecture unchanged. The corresponding base and scaled variants use the
same training data, optimization schedule, and inference configuration.
Table~\ref{tab:capacity_config} summarizes their module depths and parameter
counts.

\begin{table}[t]
  \centering
  \caption{Model configurations for the module-wise capacity study. Parameter
  counts include the TTS model only.}
  \label{tab:capacity_config}
  \vspace{0.5em}
  \small
  \setlength{\tabcolsep}{4pt}
  \begin{tabular}{lccc}
    \toprule
    Model & Causal backbone & Diff. head & Params. \\
    & layers & layers & (M) \\
    \midrule\midrule
    $\mathrm{TTS}_{50,1}$ & 7 & 4 & 228.6 \\
    \midrule
    \quad Causal backbone $\times2$ & 14 & 4 & 338.8 \\
    \quad Diffusion head $\times2$ & 7 & 8 & 294.7 \\
    \midrule\midrule
    $\mathrm{TTS}_{50,8}$ & 7 & 4 & 228.6 \\
    \midrule
    \quad Causal backbone $\times2$ & 14 & 4 & 338.8 \\
    \quad Diffusion head $\times2$ & 7 & 8 & 294.7 \\
    \bottomrule
  \end{tabular}
\end{table}

The models are trained end-to-end for 500K steps with a global batch size of up
to 40,960 packed tokens. We use the same conditional flow-matching
setup as for \modelbase{}, including four independently noised copies of each
target patch. AdamW is used with a peak learning rate of
$5\times10^{-4}$, $(\beta_1,\beta_2)=(0.9,0.95)$, and weight decay of 0.01.
The learning rate follows a cosine schedule with 5K warmup steps. To enable
CFG at inference, conditional inputs are dropped with probability 0.1 during
training.

\section{Subjective Evaluation Details}
\label{app:subjective_evaluation_details}

\begin{figure*}[t]
  \vspace{0.6em}
  \centering
  \includegraphics[width=0.9\textwidth]{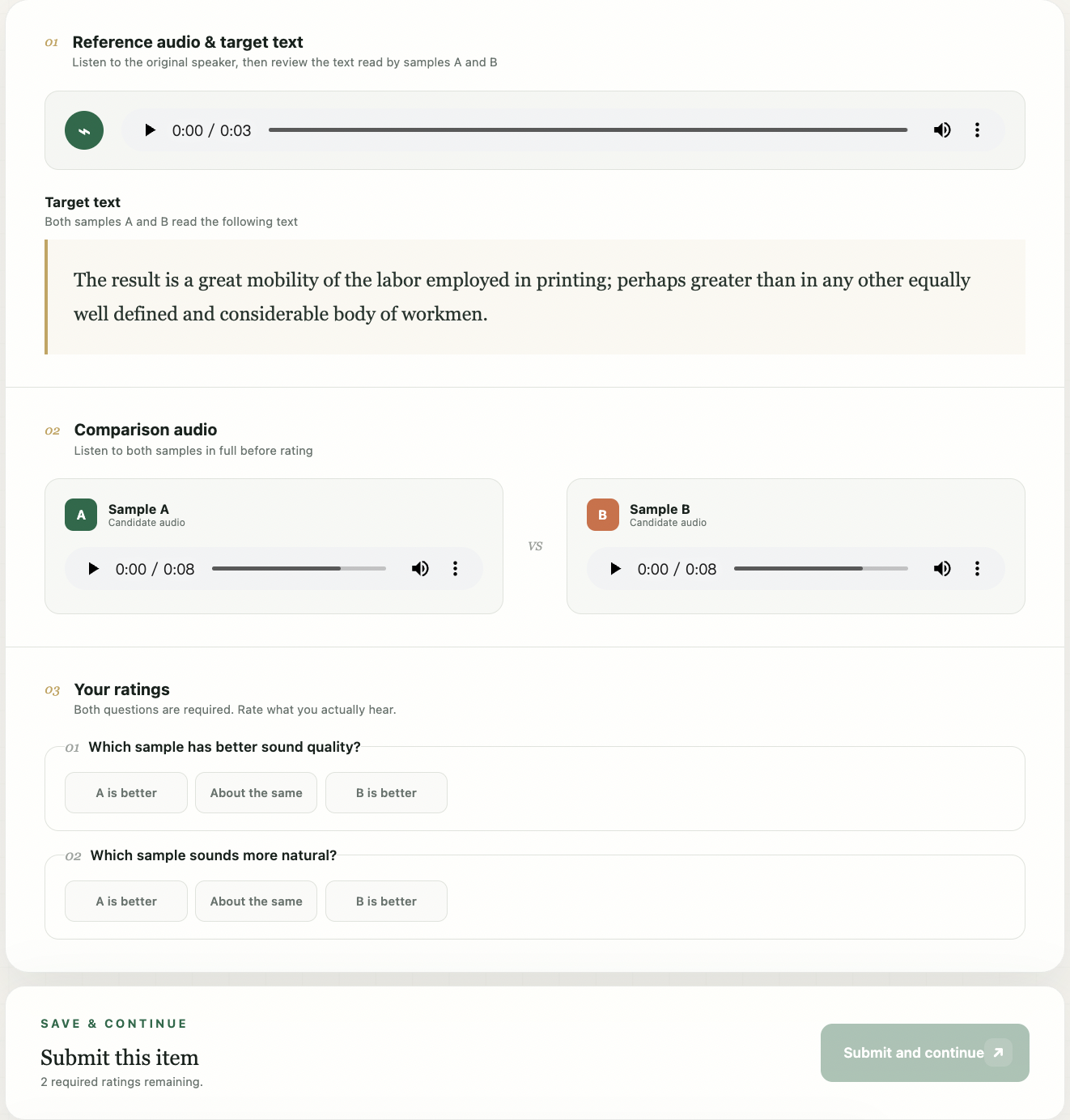}
  \caption{Annotation interface for the subjective evaluation.}
  \label{fig:subjective_annotation_interface}
\end{figure*}

We conduct the subjective evaluation on the 50-example LibriSpeech test-clean
subset.  Each of the two \cutetts{} variants is compared with MOSS-TTS-Nano,
F5-TTS, ZipVoice, and Pocket TTS on all 50 examples, yielding 400 comparison
items in total.  The items are randomly shuffled and distributed among seven
annotators.  For each pair of generated samples, annotators select the better
sample independently for sound quality and naturalness, with a tie permitted
for either dimension.  System identities are hidden throughout the evaluation.
Figure~\ref{fig:subjective_annotation_interface} shows the annotation interface.

\section{Efficiency Benchmark}
\label{app:efficiency_benchmark}

All systems are evaluated on the fixed 50-request LibriSpeech test-clean subset
using a single NVIDIA RTX~4090.  We keep one model resident at a time and
process requests strictly sequentially with batch size one.  Measurements
follow a warm-service, warm-file-cache protocol: each model completes one full
warm-up pass before measurement, while model loading, compilation, and warm-up
are excluded from the timed interval.

Each voice-cloning request begins with raw target text and a raw reference-audio
path.  Timing starts before text processing and reference file access and
includes, where applicable, audio loading and resampling, reference and speaker
encoding, LM prefill, latent generation, causal VAE decoding, and device-to-host
transfer.  No request-level tokenizer output, decoded reference audio, reference
representation, speaker embedding, or prefix state is reused across requests;
only the operating-system file cache is retained.
Table~\ref{tab:streaming_efficiency} reports the arithmetic mean, P50, and P95
first-audio latency, together with the mean per-utterance RTF, over the fixed
50 requests.

For \cutetts{}, the inference configurations follow
Appendix~\ref{app:training_inference_details}.  The base and distilled models
use the same request order and per-item seeds.  Their fixed-shape diffusion
heads and Euler sampling loops are compiled, while CUDA graphs are disabled.
Both models successfully complete all 50 requests without truncation.

For baseline inference configurations, MOSS-TTS and Qwen3-TTS use the official
vLLM-Omni inference engine, while ZipVoice uses the official
PyTriton/TensorRT engine.  The MOSS-TTS Delay model accepts a streaming request
but returns a single final PCM block; we therefore report only its RTF.
MOSS-TTS-Nano is evaluated in raw-reference voice-cloning mode.  Since
VibeVoice-Realtime does not support arbitrary-reference voice cloning, it is
evaluated using its default voice.

\end{document}